\documentclass[twocolumn,showpacs,showkeys,preprintnumbers,amsmath,amssymb,aps,nofootXinbib,floatfix]{revtex4}
\usepackage{color}
\usepackage{indentfirst}
\usepackage{graphicx}
\usepackage{multirow}
      \def\bS{{\bf S}}
      \def\bl{{\bf l}}
      \def\bp{{\bf p}}
      
      \def\br{{\bf r}}

      \def\F{{\cal F}}
      \def\H{{\cal H}}

      \def\L{{\cal L}}
      
      \def\P{{\cal P}}
      \def\R{{\cal R}}

      \def\eq{{\rm eq}}

\begin{document}

\title{Scissors mode in transuranium elements}

\author{ E.B. Balbutsev\email{balbuts@theor.jinr.ru}, I.V. Molodtsova\email{molod@theor.jinr.ru} }
\affiliation{Bogoliubov Laboratory of Theoretical Physics, Joint Institute for Nuclear Research, 141980 Dubna, Russia}

\begin{abstract}
The scissors mode is investigated in the actinides region, including even-even superheavy nuclei up to $^{256}$No,  
within the Time Dependent Hartree-Fock-Bogoliubov (TDHFB) approach.
The solution of TDHFB equations by the Wigner Function Moments (WFM) method
predicts a splitting of the scissors mode into three intermingled branches due to spin degrees of 
freedom. Both the calculated energy centroid and integrated $M1$ strength in $^{254}$No are in good agreement 
with the results of recent measurements performed by the Oslo method.
The energy centroids and summed $B(M1)$ values for others transuranium nuclides are predicted.
The calculations are performed also for $^{232}$Th and $^{236,238}$U isotopes using an updated compilation of deformation parameters.
The results are compared with that obtained previously by WFM theory and with the latest experimental data.
Progress has been achieved in theoretical understanding of the origins of double-humped structure of  scissors spectrum
observed in the Actinides.
\end{abstract}

\pacs{ 21.10.Hw, 21.60.Ev, 21.60.Jz, 24.30.Cz } 
\keywords{spin; collective motion; scissors mode; transuranium nuclides}

\maketitle

\section{Introduction}\label{I}

The theoretical interpretation of collective nuclear dynamics, realized through electromagnetic transitions 
in the low-energy region, is one of the most interesting topics in nuclear structure physics. 
A significant point in the study of the nature of collective vibrations was the idea of orbital scissors,
first reported by R. Hilton in 1976 at the conference on nuclear structure in Dubna~\cite{Hilt}. 
This idea was subsequently developed in the works of Suzuki and Rowe~\cite{Suzuki}, Lo~Iudice and Palumbo~\cite{LoIP78},
Iachello~\cite{Iachello} and other authors.

The first experimental detection of the nuclear scissors mode (SM) in  inelastic  electron  scattering on $^{156}$Gd 
at Darmstadt~\cite{Bohle} has initiated a cascade of intense experimental 
and theoretical investigations. An exhaustive review of the subject
is given in the paper~\cite{Heyd} containing about 400 references.
The nuclear SM has been extensively studied mainly by the nuclear resonance fluorescence (NRF) experiments, 
see~\cite{Heyd,NRF} and references therein. It was found that scissors excitations are a rather general phenomenon for deformed nuclei.
In even-even  rare-earth  nuclei the SM manifests itself as a concentration of magnetic dipole ($M1$) strength 
distributed over a few $1^+$ states at excitation energies between 2 and 4 MeV. 
As a rule, for strongly deformed nuclei of rare-earth elements, the measured by NRF summed strength amounts $B(M1)\simeq 3\mu_N^2$.
However, the photo-neutron measurements in $^{160-164}$Dy isotops recently performed by the Oslo group give a larger value:
$B(M1)\simeq 5\mu_N^2$~\cite{Renstrom}.

Much experimental efforts were also focused on investigation of the SM in the actinide mass region. 
Experimental studies of scissors mode excitation by NRF have  been  reported for 
$^{232}$Th~\cite{Adekola,Heil}, $^{236}$U~\cite{U236} and $^{238}$U~\cite{Heil,Hammo} isotopes.
Studies  of  the  scissors resonance excited by deuteron and $^3$He-induced reactions on $^{232}$Th in residual nuclei $^{231,232,233}$Th 
and $^{232,233}$Pa using the Oslo method have been reported in~\cite{Oslo}.The Oslo method also was applied to the $^{238}$Np nucleus~\cite{Np}.
Experimental data indicate that the strength of the SM in Actinides is larger than in Rare Earth and shifts towards lower energies.
In addition, it was found that the SM spectrum of many nuclei exhibits a distinct double-humped structure, 
see~\cite{Adekola,Oslo,Np}.

The Wigner Function Moments (WFM) or phase space moments method 
turned out to be a very convenient for the explanation of all qualitative features of the scissors mode,
see~\cite{BaSc,Ann,Malov,Malov1,BaMo,BaMoPRC13,BaMoPRC2,BaMoPRC1,BaMoPRC22,EPJ23}.
In the paper \cite{BaMo} the WFM method was applied to solve the 
time dependent Hartree-Fock equations including spin
dynamics. The most remarkable result was the prediction of a new type
of nuclear collective motion: rotational oscillations of "spin-up"
nucleons with respect to "spin-down" nucleons (the spin scissors mode). 
Further generalization which takes into account spin degrees
of freedom and pair correlations simultaneously was outlined in 
\cite{BaMoPRC2}, where the Time Dependent Hartree-Fock-Bogoliubov (TDHFB)
equations were considered.
Furthermore, after taking into account the isovector-isoscalar coupling
\cite{BaMoPRC22} one more magnetic mode (third type of scissors) emerged.
The possible existence of three scissors motions is 
easily explained by  combinatoric
consideration -- there are only three ways to divide the four different kinds 
of objects (spin up and spin down protons and neutrons in our case) into two pairs. 
The three types  of scissors modes can be approximately classified as isovector 
spin-scalar (conventional), isovector spin-vector and
isoscalar spin-vector.
In~\cite{BaMoPRC22}, the systematic calculations 
of the SM energies and $M1$ strength distributions were performed
by the WFM method and within the microscopic quasiparticle-phonon nuclear model~\cite{QPNM1,QPNM2}
 for nucleai of $N=82-126$ mass region and Actinides. 
The results of calculations were compared with the observed data obtained by the NRF experiments
and measurements performed by the Oslo method (see~\cite{BaMoPRC22} and references there).  
A remarkable coherence of both theoretical methods 
together with experimental data was observed.
In particular, the apparent splitting of the scissors spectrum in Actinides has found  a natural explanation~\cite{BaMoPRC22}.

In recent years, research of Actinides has been on the rise with the development of sensitive and sophisticated spectroscopic methods~\cite{Atoms22}.
The low availability and radioactive nature of most actinide elements make it difficult to study them experimentally. 
On this background, theoretical estimates become a particularly useful aid in the analysis of experimental results.
Among the heaviest Actinides, Fermium and Nobelium are currently the most studied, see~\cite{Spectroscopy21,Spectroscopy23} and references there.
The first experimental evidence of the SM in the superheavy nucleus $^{254}$No 
produced in the $^{208}$Pb($^{48}$Ca, $2n\gamma$)$^{254}$No reaction was recently reported~\cite{254No_Oslo}.

The present paper is an extension of our previous investigations. 
Its aim is to perform the proper WFM based calculations for even-even nuclei of transuranium elements up to $^{256}$No.
We also compare the new calculation results for $^{232}$Th and $^{236,238}$U isotopes with those obtained earlier and with experimental data.

The paper is organized as follows. We briefly describe the formalism of the WFM method in Section~\ref{II}.
The TDHFB equations for the $2\times 2$ normal and anomalous density matrices are formulated, the model Hamiltonian is presented and 
collective variables are defined. The derivation of the corresponding dynamic equations is discussed.  
Section~\ref{III} presents the calculated energies, magnetic dipole and electric quadrupole strengths of $1^+$ excitations in Actinides,
with special attention to the $^{254}$No.
The summary of main results is given in the Conclusion section~\ref{IV}.
Useful details written out in Appendix~\ref{AppA}.

\section{TDHFB equations and WFM equations of motion}\label{II}

\hspace{5mm} The TDHFB equations in matrix formulation \cite{Solov,Ring} are
\begin{equation}
i\hbar\dot\R=[\H,\R]
\label{tHFB}
\end{equation}
with
\begin{equation}
\R={\hat\rho\qquad-\hat\kappa\choose-\hat\kappa^{\dagger}\;\;1-\hat\rho^*},
\quad\H={\hat
h\quad\;\;\hat\Delta\choose\hat\Delta^{\dagger}\quad-\hat h^*}
\end{equation}
The normal density matrix $\hat \rho$ and Hamiltonian $\hat h$ are
hermitian whereas the anomalous density $\hat\kappa$ and the pairing
gap $\hat\Delta$ are skew symmetric: $\hat\kappa^{\dagger}=-\hat\kappa^*$, 
$\hat\Delta^{\dagger}=-\hat\Delta^*$.

 The microscopic Hamiltonian of the model, harmonic oscillator with 
spin orbit potential plus separable quadrupole-quadrupole and 
spin-spin residual interactions is given by
\begin{eqnarray}
\label{Ham}
 \hat h=\sum\limits_{i=1}^A\left[\frac{\hat\bp_i^2}{2m}+\frac{1}{2}m\omega^2\br_i^2
-\eta\hat \bl_i\hat \bS_i\right]+H_{qq}+H_{ss},
\end{eqnarray}
with
\begin{eqnarray}
 H_{qq}&=&\!\!\!
\sum_{\mu=-2}^{2}(-1)^{\mu}
\left\{\bar{\kappa}
 \sum\limits_i^Z\!\sum\limits_j^N
+\frac{\kappa}{2}
\left[\sum\limits_{i,j(i\neq j)}^{Z}
+\sum\limits_{i,j(i\neq j)}^{N}
\right]
\right\}
\nonumber\\&\times&\!\!\!
q_{2-\mu}(\br_i)q_{2\mu}(\br_j)
,\label{Hqq}
\\
H_{ss}&=&\!\!\!
\sum_{\mu=-1}^{1}(-1)^{\mu}
\left\{\bar{\chi}
 \sum\limits_i^Z\!\sum\limits_j^N
+\frac{\chi}{2}
\left[
\sum\limits_{i,j(i\neq j)}^{Z}
+\sum\limits_{i,j(i\neq j)}^{N}
\right]
\right\}
\nonumber\\&\times&\!\!\!
\hat S_{-\mu}(i)\hat S_{\mu}(j)
\,\delta(\br_i-\br_j),
\label{Hss}
\end{eqnarray}
where  
$\displaystyle q_{2\mu}(\br)=\sqrt{16\pi/5}\,r^2Y_{2\mu}(\theta,\phi)=
\sqrt{6}\{r\otimes r\}_{2\mu}$,
$\{r\otimes r\}_{\lambda\mu}=\sum_{\sigma,\nu}
C_{1\sigma,1\nu}^{\lambda\mu}r_{\sigma}r_{\nu}$.
Clebsch-Gordan coefficients $C_{1\sigma,1\nu}^{\lambda\mu}$,
cyclic coordinates $r_{-1}, r_0, r_1$ and spin-1/2 matrices $\hat S_{\mu}$ are defined in~\cite{Var}.
$N$ and $Z$ are numbers of neutrons and protons respectively,
$\kappa$, $\bar{\kappa}$ and $\chi$, $\bar{\chi}$ are strength constants. 
The mean field generated by this Hamiltonian was derived in \cite{BaMoPRC13}.

With the help of Fourier (Wigner) transformation the equations (\ref{tHFB}) 
are transformed into TDHFB equations for functions $f^{\tau\varsigma}(\br,
\bp,t)$ and $\kappa^{\tau\varsigma}(\br,\bp,t)$ (Wigner transforms of 
$\hat\rho$ and $\hat\kappa$) \cite{Malov,Malov1}, where $\tau$ is the isospin index,
$\varsigma\!=+,\,-,\,\uparrow\downarrow,\,\downarrow\uparrow$ is the spin index, 
  $f^{+}=f^{\upuparrows}+ f^{\downdownarrows}$ is spin-scalar function and 
  $f^{-}=f^{\upuparrows}- f^{\downdownarrows}$ is spin-vector one.
Integrating these equations over the phase space with the weights 
$\{r\otimes p\}_{\lambda\mu},\,\{r\otimes r\}_{\lambda\mu},\,
\{p\otimes p\}_{\lambda\mu}$ and  1 
one gets dynamical equations for the following
collective variables:
\begin{eqnarray}
&&L^{\tau\varsigma}_{\lambda\mu}(t)=\int\! d(\bp,\br) \{r\otimes p\}_{\lambda\mu}
 f^{\tau\varsigma}(\br,\bp,t),
\nonumber\\
&&R^{\tau\varsigma}_{\lambda\mu}(t)=\int\! d(\bp,\br) \{r\otimes r\}_{\lambda\mu}
 f^{\tau\varsigma}(\br,\bp,t),
\nonumber\\
&&P^{\tau\varsigma}_{\lambda\mu}(t)=\int\! d(\bp,\br) \{p\otimes p\}_{\lambda\mu}
 f^{\tau\varsigma}(\br,\bp,t),
\nonumber\\
&&F^{\tau\varsigma}(t)=\int\! d(\bp,\br)
 f^{\tau\varsigma}(\br,\bp,t),
\nonumber\\
&&\tilde{L}^{\tau\varsigma}_{\lambda\mu}(t)=\int\! d(\bp,\br) \{r\otimes p\}_{\lambda\mu}
 \kappa^{\tau\varsigma}(\br,\bp,t),
\nonumber\\&&
\tilde{R}^{\tau\varsigma}_{\lambda\mu}(t)=\int\! d(\bp,\br) \{r\otimes r\}_{\lambda\mu}
 \kappa^{\tau\varsigma}(\br,\bp,t),
\nonumber\\
&&\tilde{P}^{\tau\varsigma}_{\lambda\mu}(t)=\int\! d(\bp,\br) \{p\otimes p\}_{\lambda\mu}
 \kappa^{\tau\varsigma}(\br,\bp,t),
\label{Varis}
\end{eqnarray} 
where $\int\! d(\bp,\br)\equiv (2\pi\hbar)^{-3}\int\! d\br\int\! d\bp$.
The integration yields the sets of coupled (due to neutron-proton interaction)
equations for neutron and proton variables. 
The found equations are nonliner due to quadrupole-quadrupole and spin-spin
interactions. The small amplitude approximation 
allows one to linearize the equations. Writing all variables~(\ref{Varis}) as a sum of their equilibrium value plus a small deviation
\begin{eqnarray}
&&L^{\tau\varsigma}_{\lambda\mu}(t)=L^{\tau\varsigma}_{\lambda\mu}(\eq)+\L^{\tau\varsigma}_{\lambda\mu}(t),\nonumber\\
&&R^{\tau\varsigma}_{\lambda\mu}(t)=R^{\tau\varsigma}_{\lambda\mu}(\eq)+\R^{\tau\varsigma}_{\lambda\mu}(t),\nonumber\\
&&P^{\tau\varsigma}_{\lambda\mu}(t)=P^{\tau\varsigma}_{\lambda\mu}(\eq)+\P^{\tau\varsigma}_{\lambda\mu}(t),\nonumber\\
&&F^{\tau\varsigma}(t)=F^{\tau\varsigma}(\eq)+\F^{\tau\varsigma}(t),\ \ldots\nonumber
\end{eqnarray}
and neglecting quadratic deviations, one obtains the linearized equations.
It is convenient to rewrite resulting equations
in terms of isoscalar and isovector variables
$\R_{\lambda\mu}=\R_{\lambda\mu}^{\rm n}+\R_{\lambda\mu}^{\rm p}$,
$\bar \R_{\lambda\mu}=\R_{\lambda\mu}^{\rm n}-\R_{\lambda\mu}^{\rm p}$,
and so on. We also define isovector and isoscalar strength constants
$\kappa_1=\frac{1}{2}(\kappa-\bar\kappa)$ and
$\kappa_0=\frac{1}{2}(\kappa+\bar\kappa)$ connected by the relation
$\kappa_1=\alpha\kappa_0$ with $\alpha=-2$~\cite{BaSc}.
The sets of coupled dynamic equations for isovector and isoscalar variables
are written out in the Appendix~A of the Ref.~\cite{BaMoPRC22}. We are interested in the scissors mode with quantum number $K^\pi=1^+$. 
Therefore, we only need the part of dynamic equations with $\mu=1$.
As a result, we have 44 coupled isovector and isoscalar equations of the first order in
time, which can be reduced to 22 equations of the second order in time.
Excluding the integrals of motion we obtain 14 eigenvalue solutions.


\section{Results of calculations}\label{III}

The calculations are performed by the WFM method for even-even $^{226-234}$Th, $^{230-238}$U isotopes
and  nuclei of the transuranium mass region up to $^{256}$No.
The procedure of calculations and parameters values are mostly the 
same as in our previous papers \cite{BaMoPRC2,BaMoPRC1,BaMoPRC22,EPJ23}.
Particular attention is paid to the analysis of $^{254}$No, 
since experimental data on the SM in this nucleus have recently become available~\cite{254No_Oslo}.

\subsection{Transuranium nuclides}\label{IIIa}

The solutions of WFM equations for $^{254}$No are presented in the 
Table~\ref{tab1}, where the  energies of $1^+$ levels with their magnetic dipole and 
electric quadrupole strengths (see Appendix~B in~\cite{BaMoPRC22}) are shown.
It is seen that the general picture is quite similar to that one in rare-earth nuclei and the lightest Actinides~\cite{BaMoPRC22}. 

\begin{table}[h!]
\caption{The results of WFM calculations for $_{102}^{254}$No: energies $E$, magnetic dipole $B(M1)$
and electric quadrupole $B(E2)$ strengths of $1^+$ excitations. The deformation parameter $\beta_2=0.27$ ($\delta=0.255$).}
\begin{ruledtabular}\begin{tabular}{ccc}
       $E$(MeV)   & $B(M1)$($\mu_N^2$)    & $B(E2)$(W.u.)  \\    
 \hline 
   ~1.13 &  --  & 34.66  \\ 
   ~1.74 & 3.00 & ~5.57  \\ 
   ~2.50 & 5.09 & ~1.30  \\ 
   ~3.04 & 2.41 & ~6.00  \\
   ~9.51 & 0.06 & 68.11  \\ 
   11.50 & 0.00 & ~2.90  \\
   13.52 & 0.07 & ~0.04  \\
   13.63 & 0.06 & ~2.21  \\ 
   14.71 & 0.10 & ~0.70  \\ 
   15.41 & 0.01 & ~0.88  \\
   15.68 & 0.00 & ~0.42  \\ 
   15.97 & 0.25 & ~1.63  \\ 
   17.13 & 0.18 & ~1.36  \\ 
   18.40 & 3.41 & 27.85  \\            
\end{tabular}\end{ruledtabular}
\label{tab1}
\end{table} 
Among the high-lying states the isoscalar (at the energy of $9.51$~MeV) and isovector ($E=18.40$~MeV) 
GQR are distinguished by  large $B(E2)$ values. 
The rest of high-lying states have quite small excitation probabilities and we omit them from further discussion.
The lowest $1^+$ excitation at the energy $1.13$ MeV has an electrical nature. 
It was shown in~\cite{EPJ23} that it is one of three ($K^{\pi}=0^+, 1^+, 2^+$) branches of $J^{\pi}=2^+$ 
state, which can exist in a spherical nucleus (and which splits due to 
deformation into three branches with projections $\mu=0, \pm1, \pm2$).

Three low-lying magnetic $1^+$ excitations with energies $1.74$, $2.50$ and $3.04$ MeV represent various scissors modes.
The analysis of nuclear currents performed in~\cite{BaMoPRC22} has shown that every of
these excitations is the mixture of three possible  scissors: isovector
spin-scalar (conventional), isovector spin-vector and isoscalar spin-vector.
The calculated summed strength  of three scissors states amounts $B(M1) = 10.50$ $\mu_N^2$, 
the centroid is located at the energy $\bar E=2.41$ MeV.
\begin{figure}[b!]
\includegraphics[width=\columnwidth]{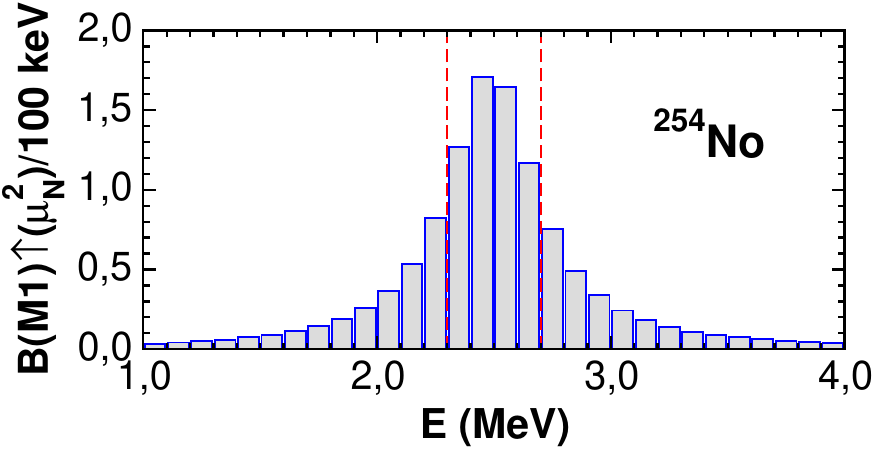}
\caption{$M1$ strength distribution in $^{254}$No from Oslo method measurements~\cite{254No_Oslo}. See text for dotted lines.}\label{fig0}
\end{figure}

Recent measurements performed by the Oslo method indicated an enhancement of $\gamma$-ray intensity in the $1.0-3.5$ MeV region
of the $^{254}$No spectrum due to the $M1$ scissors~\cite{254No_Oslo}. 
The most notaceable "bump-like" structure was observed between $1.7-3.0$ MeV with centroid at $\bar E\simeq 2.5$ MeV.
The total strength of the SM in this energy range has been estimated as $B(M1) = 11.8(19)$~$\mu_N^2$.
As can be seen from Table~\ref{tab1}, all three calculated scissors state are localized in the specified energy region.
The WFM based estimates of the summed $M1$ strength and position of the SM centroid are 
in rather good agreement with the experimental values given by the Oslo group.
More of it, the strength  distributions are quite similar. 
Really, the maximum  of the experimental  distribution represented by the gistogram in Fig.~\ref{fig0} (data from paper~\cite{254No_Oslo}) is disposed at the energy 
$\simeq 2.5$~MeV, that coincides with the calculated 
position $2.50$~MeV of the most strong low-lying excitation. The central part
of the gistogram between $2.3$~MeV and $2.7$~Mev (marked by dotted lines) contains the strength
$B(M1) \simeq 5.8\ \mu_N^2$ that is not so far from the calculated strength of
the strongest excitation $B(M1) = 5.09$~$\mu_N^2$. Two tails of the gistogram,
with $E < 2.3$ MeV and $E > 2.7$ MeV have approximately equal strengths
$B(M1) \simeq 3\ \mu_N^2$ that also agree rather well with calculated values
$B(M1) = 3.00$~$\mu_N^2$ for $E = 1.74$~MeV and $B(M1) = 2.41$~$\mu_N^2$ for $E = 3.04$~MeV.

Ref.~\cite{254No_Oslo} also presents the results of calculations of the
integrated $M1$ strength in $^{254}$No from sum rule approach and calculations 
within the quasi-particle random-phase approximation (QRPA).
The summed QRPA $M1$ strength was found to be $9.0$~$\mu_N^2$.
The calculated states are localized in the $1.8-2.9$~MeV region  with centroid at $2.4$~MeV.
Sum-rule calculations assuming a rigid-body moment of inertia~\cite{SumRule,Oslo2} (see Appendix~\ref{AppA})
predict for $^{254}$No the SM centered at $2.1$~MeV with $B(M1)=12.1(13)$~$\mu_N^2$~\cite{254No_Oslo}.
The above calculations are based on the quadrupole deformation parameter $\beta_2=0.27(3)$ reported in Refs.~\cite{No_1,No_2,No_3}.
\begin{figure}[t!]
\includegraphics[width=.9\columnwidth]{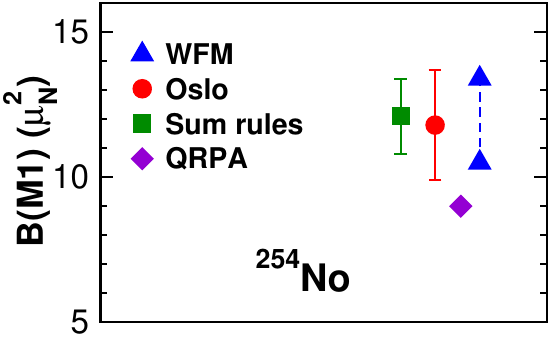}
\caption{The summed $M1$ scissors mode strength in $^{254}$No obtained from WFM calculations (blue triangles)
with the values $\beta_2=0.27$ (lower  triangle)  and $\beta_2=0.3$ (upper triangle) 
in comparison with the Oslo group measurements (red circle)
together with sum-rule estimate (green square) and QRPA calculations (violet diamond) 
reported in the Ref.~\cite{254No_Oslo}}\label{fig1}
\end{figure}
\begin{table}[t!]
\caption{
The results of WFM calculations for transuranium elements:
$E_i$ and $B_i(M1)$ are energies (in MeV) and $M1$ strengths (in $\mu_N^2$) of three ($i=1,2,3$) scissors states.
$\bar E$ is the energy centroid and $\sum B(M1)$ is summed $M1$ strength.}
\footnotesize
{
\begin{ruledtabular}\begin{tabular}{ccccccc}
  Nuclei  & $\delta$ & $i$ & $E_i$ & $B_i(M1)$   & $\bar E$ & $\sum B(M1)$\\   
\hline
                   &       & 1 &    1.77 & 3.10 &         &       \\ 
 $_{\ 94}^{238}$Pu & 0.271 & 2 &    2.58 & 5.33 &    2.52 & 11.46 \\
                   &       & 3 &    3.18 & 3.03 &         &       \\ 
                   &       & 1 &    1.76 & 3.10 &         &       \\ 
 $_{\ 94}^{240}$Pu & 0.274 & 2 &    2.61 & 5.69 &    2.55 & 11.95 \\
                   &       & 3 &    3.21 & 3.16 &         &       \\ 
                   &       & 1 &    1.76 & 3.09 &         &       \\ 
 $_{\ 94}^{242}$Pu & 0.276 & 2 &    2.64 & 6.02 &    2.58 & 12.40 \\
                   &       & 3 &    3.24 & 3.29 &         &       \\ 
                   &       & 1 &    1.74 & 3.05 &         &       \\ 
 $_{\ 94}^{244}$Pu & 0.277 & 2 &    2.66 & 6.29 &    2.60 & 12.71 \\
                   &       & 3 &    3.26 & 3.37 &         &       \\  
\hline
                   &       & 1 &    1.82 & 3.38 &         &       \\ 
 $_{\ 96}^{240}$Cm & 0.281 & 2 &    2.62 & 5.71 &    2.57 & 12.47 \\
                   &       & 3 &    3.33 & 3.38 &         &       \\ 
                   &       & 1 &    1.79 & 3.29 &         &       \\ 
 $_{\ 96}^{244}$Cm & 0.281 & 2 &    2.65 & 6.18 &    2.59 & 12.93 \\
                   &       & 3 &    3.24 & 3.46 &         &       \\ 
                   &       & 1 &    1.78 & 3.25 &         &       \\ 
 $_{\ 96}^{246}$Cm & 0.282 & 2 &    2.66 & 6.44 &    2.61 & 13.22 \\
                   &       & 3 &    3.26 & 3.54 &         &       \\ 
                   &       & 1 &    1.76 & 3.18 &         &       \\ 
 $_{\ 96}^{248}$Cm & 0.281 & 2 &    2.67 & 6.59 &    2.61 & 13.31 \\
                   &       & 3 &    3.26 & 3.55 &         &       \\
\hline
                   &       & 1 &    1.78 & 3.32 &         &       \\ 
 $_{\ 98}^{250}$Cf & 0.283 & 2 &    2.65 & 6.49 &    2.59 & 13.42 \\
                   &       & 3 &    3.24 & 3.61 &         &       \\  
                   &       & 1 &    1.78 & 3.33 &         &       \\ 
 $_{\ 98}^{252}$Cf & 0.288 & 2 &    2.69 & 6.96 &    2.64 & 14.11 \\
                   &       & 3 &    3.29 & 3.82 &         &       \\ 
\hline
                   &       & 1 &    1.81 & 3.48 &         &       \\ 
 $_{ 100}^{250}$Fm & 0.280 & 2 &    2.59 & 5.91 &    2.54 & 12.87 \\
                   &       & 3 &    3.18 & 3.48 &         &       \\ 
                   &       & 1 &    1.78 & 3.38 &         &       \\ 
 $_{ 100}^{252}$Fm & 0.277 & 2 &    2.59 & 5.98 &    2.53 & 12.78 \\
                   &       & 3 &    3.17 & 3.42 &         &       \\ 
                   &       & 1 &    1.73 & 3.20 &         &       \\ 
 $_{ 100}^{254}$Fm & 0.266 & 2 &    2.52 & 5.60 &    2.46 & 11.85 \\
                   &       & 3 &    3.10 & 3.06 &         &       \\ 
                   &       & 1 &    1.71 & 3.10 &         &       \\ 
 $_{ 100}^{256}$Fm & 0.261 & 2 &    2.50 & 5.55 &    2.43 & 11.57 \\
                   &       & 3 &    3.07 & 2.93 &         &       \\ 
\hline
                   &       & 1 &    1.84 & 3.75 &         &       \\ 
 $_{ 102}^{250}$No & 0.281 & 2 &    2.57 & 5.61 &    2.51 & 12.86 \\
                   &       & 3 &    3.15 & 3.49 &         &       \\ 
                   &       & 1 &    1.83 & 3.69 &         &       \\ 
 $_{ 102}^{252}$No & 0.281 & 2 &    2.58 & 5.85 &    2.52 & 13.07 \\
                   &       & 3 &    3.16 & 3.53 &         &       \\ 
                   &       & 1 &    1.80 & 3.60 &         &       \\ 
 $_{ 102}^{254}$No & 0.279 & 2 &    2.57 & 5.99 &    2.52 & 13.08 \\
                   &       & 3 &    3.15 & 3.49 &         &       \\ 
                   &       & 1 &    1.76 & 3.43 &         &       \\ 
 $_{ 102}^{256}$No & 0.270 & 2 &    2.52 & 5.72 &    2.46 & 12.33 \\
                   &       & 3 &    3.10 & 3.19 &         &       \\  
\end{tabular}\end{ruledtabular}
}\label{tab2}
\end{table}
\clearpage

The results of the WFM theory presented in the Table~\ref{tab1} are obtained for the same value $\beta_2=0.27$.
A recent in-beam $\gamma$ spectroscopic study of the $^{252,254}$No isotopes~\cite{252No,No_4} resulted in a revised deformation parameter $\beta_2=0.32(2)$ for $^{254}$No
and gave $\beta_2=0.31(2)$ value for $^{252}$No. 
Our calculations performed for $^{254}$No with $\beta_2=0.3$ ($\delta=0.284$) lead to an increase in the 
summed strength up to $B(M1)=13.40$~$\mu_N^2$ with the centroid at $\bar E=2.60$~MeV.

It has been experimentally established that in the  mass region covering transitional and strongly deformed rare-earth nuclei, 
the behavior of the low-lying summed $M1$ strength generally follows a quadratic dependence on deformation~\cite{Heyd}.
The strong dependence of the SM strength on deformation is also evident for Actinides~\cite{Oslo2}.
A comparison of the WFM theory with the experiment and theoretical estimates presented in the paper~\cite{254No_Oslo}
is shown in Fig.~\ref{fig1}, where the WFM results are displayed for two $\beta_2$ values ($0.27$ and $0.3$). 
As can be seen, the predictions of the WFM theory are in good agreement both 
with experiment and with other theories.

The results of our calculations for transuranium nuclei are presented in the Tabel~\ref{tab2}.
The energies $E_i$ of three ($i=1,2,3$) low-lying magnetic $1^+$ excitations (three types of SM) with the corresponding $B_i(M1)$ values 
are shown together with their energy centroids $\bar E$ and summed $M1$ strengths.
The quadrupole deformation parameter $\delta$ relates to $\beta_2$ as $\delta\simeq \beta_2 \sqrt{45/16\pi}$. 
The deformation parameters for Pu-Cf are taken from compilation of Raman~{\it et al.}~\cite{Raman} and for Fm and No -- from
the global large-scale axial relativistic Hartree-Bogoliubov calculations for even-even nuclei of Abgemava~{\it et al.}~\cite{Ring2}. 
According to our calculations, in all the nuclei studied, scissors excitations are located in the energy range from $1.7$ to $3.3$ MeV 
with the centroid $\bar E=2.5-2.6$~MeV. The total $M1$ strength amounts $\sum B(M1)=11.5-13.5$~$\mu_N^2$.
Using the value of $\beta_2=0.295$~($\delta=0.279$) given by Ref.~\cite{Ring2} for $^{254}$No, we arrive at $\sum B(M1)=13.08$~$\mu_N ^2$ 
with $\bar E=2.52$~MeV, see Table~\ref{tab2}.

\begin{table}[t!]
\caption{
The results of calculations by the WFM method. 
The notations are the same as in the Table~\ref{tab2}.
 }
{
\begin{ruledtabular}\begin{tabular}{ccccccc}
  Nuclei  & $\delta$ & $i$ & $E_i$ & $B_i(M1)$   & $\bar E$ & $\sum B(M1)$\\   
          &          &     & (MeV) & $(\mu_N^2)$ & (MeV) & $(\mu_N^2)$  \\
\hline
                   &       & 1 &    1.64 & 1.96 &         &       \\ 
 $_{\ 90}^{226}$Th & 0.216 & 2 &    2.24 & 2.15 &    2.16 & ~5.38 \\
                   &       & 3 &    2.81 & 1.26 &         &       \\ 
                   &       & 1 &    1.63 & 1.99 &         &       \\ 
 $_{\ 90}^{228}$Th & 0.218 & 2 &    2.26 & 2.39 &    2.18 & ~5.73 \\
                   &       & 3 &    2.83 & 1.35 &         &       \\ 
                   &       & 1 &    1.65 & 2.24 &         &       \\ 
 $_{\ 90}^{230}$Th & 0.231 & 2 &    2.35 & 3.16 &    2.28 & ~7.16 \\
                   &       & 3 &    2.93 & 1.76 &         &       \\ 
                   &       & 1 &    1.68 & 2.46 &         &       \\ 
 $_{\ 90}^{232}$Th & 0.247 & 2 &    2.47 & 4.04 &    2.40 & ~8.77 \\
                   &       & 3 &    3.06 & 2.27 &         &       \\ 
                   &       & 1 &    1.62 & 2.14 &         &       \\ 
 $_{\ 90}^{234}$Th & 0.228 & 2 &    2.35 & 3.32 &    2.27 & ~7.20 \\
                   &       & 3 &    2.93 & 1.75 &         &       \\ 
\hline
                   &       & 1 &    1.73 & 2.73 &         &       \\ 
 $_{\ 92}^{230}$U  & 0.248 & 2 &    2.43 & 3.70 &    2.36 & ~8.64 \\
                   &       & 3 &    3.02 & 2.21 &         &       \\ 
                   &       & 1 &    1.72 & 2.73 &         &       \\ 
 $_{\ 92}^{232}$U  & 0.250 & 2 &    2.45 & 3.98 &    2.38 & ~9.01 \\
                   &       & 3 &    3.04 & 2.30 &         &       \\ 
                   &       & 1 &    1.73 & 2.81 &         &       \\ 
 $_{\ 92}^{234}$U  & 0.257 & 2 &    2.51 & 4.55 &    2.44 & ~9.92 \\
                   &       & 3 &    3.10 & 2.57 &         &       \\ 
                   &       & 1 &    1.74 & 2.91 &         &       \\ 
 $_{\ 92}^{236}$U  & 0.267 & 2 &    2.58 & 5.25 &    2.52 & 11.07 \\
                   &       & 3 &    3.18 & 2.91 &         &       \\ 
                   &       & 1 &    1.74 & 2.92 &         &       \\ 
 $_{\ 92}^{238}$U  & 0.271 & 2 &    2.62 & 5.66 &    2.56 & 11.67 \\
                   &       & 3 &    3.22 & 3.09 &         &       \\  
\end{tabular}\end{ruledtabular}
}\label{tab3}
\end{table}

\subsection{Lightest Actinides}

The scissors excitations in $^{232}$Th and  $^{236,238}$U isotopes were initially investigated in the NRF experiments~\cite{Heyd,Adekola,Heil,U236,Hammo,SumRule}.
A duble-hump structure of the SM in $^{232}$Th was first observed by Adekola~{\it et al.}~\cite{Adekola}.
Later, the Oslo method was applied to study a number of the actinide region nuclei~\cite{Oslo,Oslo2,Np,254No_Oslo}.
Oslo experiments revealed that a similar separation of the scissors mode into two components is typical for most of the studied nuclides.
%
\begin{table*}[t!]
\caption{Scissors mode energy centroids $\bar E$ and summed strengths $\sum B(M1)$
from Oslo~\cite{Oslo2} and NRF~\cite{Adekola,U236,Hammo} experiments are compared with WFM calculations and sum rule (SR) estimates. 
The energy range of summation is indicated in the last line. For deformation parameters, see the Table~\ref{tab3}.}
\begin{ruledtabular}\begin{tabular}{lcccccccccccc}
 & & \multicolumn{5}{c}{Experiment} &  \multicolumn{6}{c}{Theory}\\   
\cline{2-7}\cline{8-13}
 & & \multicolumn{2}{c}{$^{232}$Th} &  $^{236}$U  & \multicolumn{2}{c}{$^{238}$U}  &
    \multicolumn{2}{c}{$^{232}$Th}             &  \multicolumn{2}{c}{$^{236}$U}  & \multicolumn{2}{c}{$^{238}$U}\\  
\cline{3-4}\cline{5-5} \cline{6-7}       
\cline{8-9}\cline{10-11}\cline{12-13}
 & Ref. & \cite{Oslo2} & \cite{Adekola} &  \cite{U236}  & \cite{Oslo2} & \cite{Hammo}  &
    WFM          & SR             &  WFM          & SR           & WFM & SR\\
 \hline    
$\bar E$ (MeV)           & & 2.23(14) & 2.5 & 2.3 & 2.24(15) & 2.6(6) &   2.40 & 2.06 & 2.52 & 2.20 & 2.56 & 2.22  \\
\multicolumn{2}{l}{$\sum B(M1)$ ($\mu_N^2$)}  & 9.5(26)  & 4.3(6) & 4.1(6) & 9.4(16) & 8(1) &    8.77 & 9.62 & 11.07 & 10.71 & 11.67 & 10.81 \\
Range (MeV) & & $1-4$ & $2-4$ & $1.8-3.2$ & $1-4$ & $2-4.3$ & $1.6-4$ & & $1.7-4$ & & $1.7-4$ & \\
\end{tabular}\end{ruledtabular}\label{tab4}
\end{table*}

The features of the scissor mode in $^{232}$Th and  $^{236,238}$U  were studied within WFM method in the papers~\cite{BaMoPRC1,BaMoPRC22}.
Theoretical estimates obtained in these works were compared with the results of NRF measurements. 
Thus, an explanation for the observed splitting of the SM strength was proposed. 
As has been shown, the separation may be related to three types of the SM (see~\cite{BaMoPRC1,BaMoPRC22} for details).

The calculations mentioned above were performed using the deformation parameters taken from Ref.~\cite{SumRule}.
In the present calculations, we use the deformation values from the compilation of Raman~{\it et al.}~\cite{Raman}.
The WFM results for even-even $^{226-234}$Th and $^{230-238}$U isotopes are summarized in the Table~\ref{tab3}.
The SM in $^{231-233}$Th, $^{232,233}$Pa, and $^{237-239}$U isotopes has been studied  by the Oslo method in the paper~\cite{Oslo2}.
A comparison of the WFM theory predictions for  energy centroids  $\bar E$ and the summed strengths $\sum B(M1)$ in $^{232}$Th 
and $^{236,238}$U  with the results of measurements performed by the Oslo group and obtained from the NRF experiments~\cite{Adekola,U236,Hammo}
is given in the Table~\ref{tab4}. 
The sum-rule (SR) estimates (see Appendix~\ref{AppA}) are also included in this Table.
Both the $\bar E$ and $\sum B(M1)$ values obtained within WFM theory and measured by the Oslo method are in good agreement for $^{232}$Th.
The $M1$ strength also agrees with the sum-rule estimates. However, the strength detected by NRF is half that.
This may be due to the fact that summation in NRF measurements is limited by the energy range $2-4$ MeV. 
The same reason applies to $^{236}$U.

The calculated $\sum B(M1)$ values for $^{238}$U turn out to be slightly overestimated compared to the Oslo, NRF and SR results, 
which agree with each other rather good. 
It should also be noted that our calculations were carried out keeping the same set of parameters for all actinide nuclei.
It is clear that a slight adjustment of the parameters could improve the agreement with the experiment for $^{238}$U, but 
we do not strive for this. Our interest is to reveal the general trends in SM behaviour throughout actinide mass region.

A comparison of the experimental results and the theoretical estimates compiled in the Table~\ref{tab4} 
leads to the observation that Eq.~(\ref{ESM}) of the sum rule approach produces slightly underestimated values for the SM centroid in Actinides.
The same conclusion arises from the comparison of corresponding results for $^{254}$No.
So, the results of the Oslo method measurements, the QRPA and WFM estimates for the centroid are $2.5,\ 2.4$ and $2.52$~MeV, respectively
(see previous subsection), whereas the sum-rule estimate is only $2.06$~MeV. 

\begin{figure}[h!]
\includegraphics[width=\columnwidth]{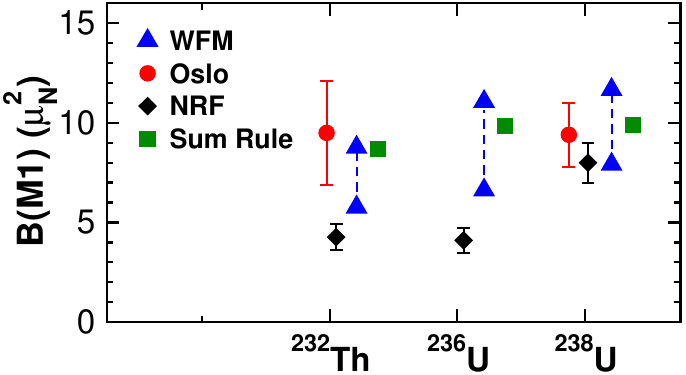}
\caption{The summed scissors mode strength 
from WFM calculations (blue triangles)
with the deformation parameters $\delta=0.216,\ 0.225,\ 0.234$ for  $^{232}$Th, $^{236}$U,  $^{238}$U, respectively (lower triangle)~\cite{BaMoPRC22}
and with the $\delta$ values listed in Table~\ref{tab3} (upper triangle) 
in comparison with the Oslo method~\cite{Oslo2} (red circles) and NRF~\cite{U236,Hammo} (black diamonds) results 
together with sum-rule estimates (green squares), see text.
}\label{fig2}
\end{figure}

The results of calculations and measurements of the SM strength, listed in the Table~\ref{tab4}, 
are also plotted in the Fig.~\ref{fig2}.
In addition, this figure shows the results of our previous calculations~\cite{BaMoPRC22} with deformation parameters taken from~\cite{SumRule}.
These results can be considered as a lower limit. 
In general, keeping in mind the scatter in the values of the deformation parameters, 
one can conclude that the results of the WFM theory are in good agreement with both the latest experimental data and the sum-rule estimates.

\begin{table*}[t!]
\caption{The energy centroids for lower $E_{\rm L}$ and upper $E_{\rm U}$ scissors mode (SM) and corresponding strengths $B_{\rm L}$, $B_{\rm U}$: 
WFM theory results are compared with the Oslo~\cite{Oslo2} experiments. 
The deformation parameters are taken from~\cite{Raman}.}
\begin{ruledtabular}\begin{tabular}{lccccccccc}
 & & \multicolumn{4}{c}{Lower SM} &  \multicolumn{4}{c}{Upper SM}\\   
\cline{3-6}\cline{7-10}
$^A$X & $\delta$ & \multicolumn{2}{c}{$E_{\rm L}$ (MeV)}  & \multicolumn{2}{c}{$B_{\rm L}$ ($\mu_N^2$)}  & 
                     \multicolumn{2}{c}{$E_{\rm U}$ (MeV)}  & \multicolumn{2}{c}{$B_{\rm U}$ ($\mu_N^2$)}\\
\cline{3-4}\cline{5-6} \cline{7-8}  \cline{9-10}       
 & & WFM & Oslo & WFM  & Oslo & WFM & Oslo & WFM & Oslo \\
 \hline    
 $^{232}$Th& 0.247 & 2.17 & 1.95(15) & 6.50 & 6.5(22) & 3.06 & 2.85(10) & 2.27& 3.0(12)  \\
 $^{238}$U & 0.271 & 2.32 & 1.95(15) & 8.58 & 6.5(12) & 3.22 & 2.90(15) & 3.09& 2.9(10)  \\
\end{tabular}\end{ruledtabular}\label{tab5}
\end{table*}

As mentioned at the beginning of this subsection, the double-bump structure of the SM has been observed in several actinide nuclei.
The splitting of the SM into two components has been reported by Oslo group for $^{231-233}$Th, $^{232,233}$Pa, $^{237-239}$U isotopes~\cite{Oslo2} 
and for odd-odd $^{238}$Np nucleus~\cite{Np}.
The WFM theory predicts a splitting of the SM into three intermingled branches~\cite{BaMo,BaMoPRC1,BaMoPRC22}. 
The energies $E_i$ ($i=1,2,3$) of three low-lying $1^+$ states (branches of SM) with the corresponding $M1$ strengths for $^{232}$Th and $^{238}$U 
are contained in the Tabel~\ref{tab3}. 
The energy centroid $E_{\rm L}$ and summed strength $B_{\rm L}$ of two lowest scissors states (with $i=1$ and $i=2$) are given by
$E_{\rm L}=\left(E_1 B_1+E_2 B_2\right)/B_{\rm L}$ and $B_{\rm L}=B_1+B_2$. 
This bunch can be compared to the observed in the experiment lower in energy group of levels.
The third state with $E_{\rm U}=E_3$ and $B_{\rm U}=B_3$ can be associated with the upper observed group.
Such comparison is presented in the Table~\ref{tab5}.
As can be seen, the agreement between theory and experiment for $^{232}$Th is excellent.
For $^{238}$U, the theory gives a slight excess in strength for the lower SM.
In general, the theory adequately reproduces the observed distribution of the $M1$ strength.

The analysis of nuclear currents carried out in paper~\cite{BaMoPRC22} revealed that 
excitations localized around 3~MeV are predominantly due to the conventional excitation mechanism SM.
Two lowest in energy states arise when spin degrees of freedom are taken into account~\cite{BaMo}.
They are predominantly a mixture of two types of spin scissors (spin-vector isoscalar and spin-vector isovector). 
Based on our analysis, it can be assumed that the observed splitting of SM occurs due to the separation of conventional scissors ($E_{\rm U}$) 
and spin-scissors  ($E_{\rm L}$) excitations.

\section{Conclusion}\label{IV}

In this work, we have extended our previous WFM investigations of the scissor mode in the rare-earth elements, 
$^{232}$Th and $^{236,238}$U isotopes~\cite{BaMo,BaMoPRC1,BaMoPRC22} to the region of the transuranium nuclei.
The trigger for this was the recent reported observation by the Oslo group of the SM in the $^{254}$No nucleus~\cite{Oslo}.
The calculated summed strength in $^{254}$No amounts $\sum B(M1) = 13.08$~$\mu_N^2$, the centroid being located at the energy $\bar E=2.52$ MeV.
These results 
are in good agreement with both the energy centroid ($\simeq 2.5$~MeV) and integrated $M1$ strength ($11.8(19)$ $\mu_N^2$) 
detected in the experiment.
Moreover, the calculated strength distribution is very similar to the observed one.

The SM energy centroids and $B(M1)$ values for the even-even transuranium nuclides up to $^{256}$No are predicted.
In all nuclei, scissors excitations are located in the energy range from $1.7$ to $3.3$~MeV 
with centroids $\bar E=2.5-2.6$~MeV. The total $M1$ strengths amount $\sum B(M1)=11.5-13.5$~$\mu_N^2$.

A double-humped structure of the SM in $^{232}$Th has been first observed in NRF experiment~\cite{Adekola}.
The splitting of the SM into two components has been reported by Oslo group for $^{231-233}$Th, $^{232,233}$Pa, 
$^{237-239}$U isotopes~\cite{Oslo2}. 
A new WFM calculations were performed for $^{232}$Th and $^{236,238}$U using an updated compilation of deformation parameters.
The WFM theory is in good agreement with the observed by the Oslo group integral characteristics and 
adequately reproduces the distribution of the $M1$ strength in $^{232}$Th and $^{238}$U.
Our analysis allows to conclude that the observed splitting of SM occurs due to the separation of conventional sissors
and spin-scissors excitations.

\begin{acknowledgments}
We are grateful to F. L. Bello Garrote for providing us with data for the Figure~\ref{fig0}.
\end{acknowledgments}

\appendix

\section{Sum rule approach}
\label{AppA}

The sum rule approach was introduced by Enders~{\it et al.}~\cite{SumRule} in order to describe the excitation energy and strength of the scissors mode 
in even-even nuclei. From a theoretical consideration it follows that the SM strength must be proportional to the moment of inertia of the nucleus, see~\cite{Heyd}. 
As it was argued in paper~\cite{Oslo2}, in the actinides region a rigid-body moment of inertia should be applied.
All the equations written below were derived in the papers~\cite{Heyd,SumRule,Oslo2}. Details can be found there.
According to the Ref.~\cite{Oslo2}, the scissors mode centroid $\omega_{\rm SM}$ and strength $B_{\rm SM}$ (in authors' notations) are given by:
\begin{eqnarray}
&&\omega_{\rm SM}=\omega_D\delta \sqrt{2\xi},\label{ESM}\\
&&B_{\rm SM}=\frac{3}{4}\left(\frac{Z}{A}\right)^2\Theta_{\rm rigid}\omega_D\delta \sqrt{2\xi},\label{BSM}
\end{eqnarray}
where
$
 \Theta_{\rm rigid}=\frac{2}{5}m_Nr_0^2A^{5/3}\left(1+0.31\delta\right)
$
is the rigid-body moment of inertia, 
$\xi=\omega_Q^2/\left(\omega_Q^2+2\omega_D^2\right)$ and
\begin{eqnarray}
&&\omega_D\approx(31.2A^{-1/3}+20.6A^{-1/6})\left(1-0.61\delta\right){\rm MeV},\hspace*{7mm}   \\
&&\omega_Q\approx64.7A^{-1/3}\left(1-0.3\delta\right){\rm MeV} \label{OmQ}
\end{eqnarray}
are the energy centroids of the isovector giant dipole and isoscalar giant quadrupole resonances, respectively.

In our calculations we use the deformation parameters taken from the compilation of Raman {\it et al.}~\cite{Raman}, the value $r_0=1.2$ fm.


\begin{thebibliography}{99}

\bibitem{Hilt}
R. R. Hilton, {\it Talk presented at the International Conference on
Nuclear Structure} (Joint Institute for Nuclear Research, Dubna, Russia, 1976) (unpublished).

\bibitem{Suzuki}
T. Suzuki and D. J. Rowe, Nucl. Phys. A {\bf 289} (1977) 461.

\bibitem{LoIP78}
N. Lo Iudice and F. Palumbo, Phys. Rev. Lett. {\bf 41} (1978) 1532.

\bibitem{Iachello}
F. Iachello, Nucl. Phys. A {\bf 358}, 89c (1981).

\bibitem{Bohle}
D. Bohle, A. Richter, W. Steffen, A. E. L. Dieperink, N. Lo Iudice, 
F. Palumbo, and O. Scholten, Phys. Lett. B {\bf 137}, 27 (1984).


\bibitem{Heyd} 
K. Heyde, P. von Neumann-Cosel, and A. Richter,
Rev. Mod. Phys. {\bf 82}, 2365 (2010).

\bibitem{NRF}
U. Kneissl, H. H. Pitz and A. Zilges, Prog. Part. Nucl. Phys. {\bf 37}, 349 (1996).

\bibitem{Renstrom}
T. Renstr\o{}m, H. Utsunomiya, H. T. Nyhus, A. C. Larsen, M. Guttormsen, G. M. Tveten, 
D. M. Filipescu, I. Gheorghe, S. Goriely, S. Hilaire, Y.-W. Lui, J. E. Midtb\o{}, S. P\'{e}ru, T. Shima, S. Siem, and
O.~Tesileanu, Phys.~Rev.~C~{\bf 98}, 054310 (2018).

\bibitem{Adekola}
A. S. Adekola, C. T. Angell, S. L. Hammond, A. Hill, C. R. Howell, H. J. Karwowski, 
J. H. Kelley, and E. Kwan, Phys. Rev. C {\bf 83}, 034615 (2011).

\bibitem{Heil}
R. D. Heil, H. H. Pitz, U. E. P. Berg, U. Kneissl, K. D. Hummel, G. Kilgus, D. Bohle, A. Richter,  C. Wesselborg,  and  P. von Brentano,
Nucl. Phys. A {\bf 476}, 39 (1988).

\bibitem{U236}
J. Margraf, A. Degener, H. Friedrichs, R. D. Heil, A. Jung, U. Kneissl,
S. Lindenstruth, H. H. Pitz, H. Schacht, U. Seemann, R. Stock, C. Wesselborg,
P. von Brentano, and A. Zilges,
Phys. Rev. C {\bf 42}, 771 (1990).

\bibitem{Hammo}
S. L. Hammond, A. S. Adekola, C. T. Angell, H. J. Karwowski, E. Kwan, G. Rusev,
A.P. Tonchev, W. Tornow, C. R. Howell, and J. H. Kelley, Phys. Rev. C {\bf 85}, 044302 (2012).

\bibitem{Oslo}
M. Guttormsen, L. A. Bernstein, A. B\"{u}rger, A. G\"{o}rgen, F. Gunsing, T. W. Hagen, A. C. Larsen, T. Renstr\o{}m, S. Siem,
M. Wiedeking, and J. N. Wilson, Phys. Rev. Lett. {\bf 109}, 162503 (2012).

\bibitem{Np} 
T. G. Tornyi, M. Guttormsen, T. K. Eriksen, A. G\"{o}rgen, F. Giacoppo, T. W. Hagen, A. Krasznahorkay, A. C. Larsen, 
T. Renstr\o{}m, S. J. Rose, S. Siem, and G. M. Tveten, Phys. Rev. C {\bf 89}, 044323 (2014).


\bibitem{BaSc}
E. B. Balbutsev and P. Schuck, Nucl. Phys. A {\bf 720}, 293 (2003); {\bf 728}, 471 (2003).

\bibitem{Ann}
E. B. Balbutsev and P. Schuck, Ann. Phys. {\bf 322}, 489  (2007).

\bibitem{Malov}
E. B. Balbutsev, L. A. Malov, P. Schuck, M. Urban, X. Vi\~nas, Phys. At. Nucl. {\bf 71}, 1012-1030 (2008).

\bibitem{Malov1}
E. B. Balbutsev, L. A. Malov, P. Schuck, M. Urban, Phys. Atom. Nuclei {\bf 72}, 1305-1319 (2009).

\bibitem{BaMo}
E. B. Balbutsev, I.V. Molodtsova, and P. Schuck, Nucl. Phys. A {\bf 872}, 42 (2011).

\bibitem{BaMoPRC13}
E. B. Balbutsev, I.V. Molodtsova, and P. Schuck, Phys. Rev. C {\bf 88}, 014306 (2013).

\bibitem{BaMoPRC2}
E. B. Balbutsev, I.V. Molodtsova, and P. Schuck, Phys. Rev. C {\bf 91}, 064312 (2015).

\bibitem{BaMoPRC1}
E. B. Balbutsev, I.V. Molodtsova, and P. Schuck, Phys. Rev. C {\bf 97},  044316 (2018).

\bibitem{BaMoPRC22}
E. B. Balbutsev, I.V. Molodtsova, A. V. Sushkov,  N. Yu. Shirikova, and P. Schuck, Phys. Rev. C {\bf 105},  044323 (2022).

\bibitem{EPJ23} E. B. Balbutsev and I.V. Molodtsova, Eur. Phys. J. A {\bf 59}, 207 (2023).


\bibitem{QPNM1} V. G. Soloviev, A. V. Sushkov, N. Yu. Shirikova, and N. LoIudice, Nucl. Phys. A {\bf 600}, 155 (1996).
\bibitem{QPNM2} V. G. Soloviev, A. V. Sushkov, and N. Yu. Shirikova, Phys. Part. Nucl. {\bf 31}, 385 (2000).


\bibitem{Atoms22}
M. Laatiaoui and S. Raeder, Atoms {\bf10}, 61 (2022).
\bibitem{Spectroscopy21}
M. Block, M. Laatiaoui and S. Raeder,
Prog. Part. Nucl. Phys. {\bf 116}, 103834 (2021).
\bibitem{Spectroscopy23}
X. F. Yang, S. J. Wang, S. G. Wilkins and R. F. Garcia Ruiz
Prog. Part. Nucl. Phys. {\bf 129}, 104005 (2023).

\bibitem{254No_Oslo}
F. L. Bello Garrote, A. Lopez-Martens, A.C. Larsen, I. Deloncle, S. P\'{e}ru,
F. Zeiser, P.T. Greenlees, B.V. Kheswa, K. Auranen, D.L. Bleuel, D.M. Cox,
L. Crespo Campo, F. Giacoppo, A. G\"{o}rgen, T. Grahn, M. Guttormsen, T.W. Hagen,
L. Harkness-Brennan, K. Hauschild, G. Henning, R.-D. Herzberg, R. Julin,
S. Juutinen, T.A. Laplace, M. Leino, J.E. Midtb\o{}, V. Modamio, J. Pakarinen,
P. Papadakis, J. Partanen, T. Renstr\o{}m, K. Rezynkina, M. Sandzelius, J. Sar\'{e}n,
C. Scholey, S. Siem, J. Sorri, S. Stolze, and J. Uusitalo, Phys. Lett. B {\bf 834}, 137479 (2022).

\bibitem{Solov}
V. G. Soloviev, {\it Theory of complex nuclei} (Pergamon Press, Oxford, 1976). 

\bibitem{Ring} P. Ring and P. Schuck,
 {\it The Nuclear Many-Body Problem} (Springer, Berlin, 1980).

\bibitem{Var}
 D. A. Varshalovitch, A. N. Moskalev, and V. K. Khersonski,
{\it Quantum Theory of Angular Momentum} (World Scientific, Singapore, 1988).


 \bibitem{No_1}
 P. Reiter, T. L. Khoo, C. J. Lister, D. Seweryniak, I. Ahmad, M. Alcorta, M. P. Carpenter, 
 J. A. Cizewski, C. N. Davids, G. Gervais, J. P. Greene, W. F. Henning, R. V. F. Janssens, T. Lauritsen, S. Siem, A. A. Sonzogni, 
 D. Sullivan, J. Uusitalo, I. Wiedenh\"{o}ver, N. Amzal, P. A. Butler, A. J. Chewter, K. Y. Ding, N. Fotiades, J. D. Fox, P. T. Greenlees, R.-D. Herzberg, 
 G. D. Jones, W. Korten, M. Leino, and K. Vetter,  Phys. Rev. Lett. {\bf 82}, 509 (1999). 
\bibitem{No_2}  
 M. Leino {\it et al.}, Eur. Phys. J. A {\bf 6},63 (1999).
\bibitem{No_3} 
 S. Raeder, D. Ackermann, H. Backe, R. Beerwerth, J. C. Berengut, M. Block, {\it et al.}, Phys. Rev. Lett. {\bf 120}, 232503 (2018).
 
\bibitem{SumRule}  
J. Enders, P. von Neumann-Cosel, C. Rangacharyulu, and A. Richter
Phys. Rev. C {\bf 71}, 014306 (2005).

\bibitem{Oslo2} 
 M. Guttormsen, L. A. Bernstein, A. G\"{o}rgen, B. Jurado, S. Siem, M. Aiche, Q. Ducasse, F. Giacoppo, F. Gunsing, T. W. Hagen, 
 A. C. Larsen, M. Lebois, B. Leniau, T. Renstr\o{}m, S. J. Rose, T. G. Tornyi, G. M. Tveten, M. Wiedeking, and J. N. Wilson,
 Phys. Rev. C {\bf 89}, 014302 (2014).
 
\bibitem{252No}
 R.-D. Herzberg, N. Amzal, F. Becker, P. A. Butler, A. J. C. Chewter, J. F. C. Cocks, 
 O. Dorvaux, K. Eskola, J. Gerl, P. T. Greenlees, N. J. Hammond, K. Hauschild, K. Helariutta, F. Hessberger, M. Houry, 
 G. D. Jones, P. M. Jones, R. Julin, S. Juutinen, H. Kankaanpaa, H. Kettunen, T. L. Khoo, W. Korten, P. Kuusiniemi, 
 Y. LeCoz, M. Leino, C. J. Lister, R. Lucas, M. Muikku, P. Nieminen, R. D. Page, P. Rahkila, P. Reiter, 
 C. Schlegel, C. Scholey, O. Stezowski, C. Theisen, W. H. Trzaska, J. Uusitalo, and H. J. Wollersheim, Phys. Rev. C {\bf 65}, 014303 (2001). 
 
\bibitem{No_4}
 P. A. Butler, R. D. Humphreys, P. T. Greenlees, R. D. Herzberg, D. G. Jenkins, G. D. Jones, {\it et al.}, Phys. Rev. Lett. {\bf 89}, 202501 (2002). 
 
\bibitem{Raman} 
S. Raman, C. W. Nestor, and P. Tikkanen,
At. Data Nucl. Data Tables {\bf 78}, 1 (2001).
 
\bibitem{Ring2} 
 S. E. Agbemava, A. V. Afanasjev, D. Ray, and P.~Ring, Phys. Rev. C {\bf 89}, 054320 (2014).
 
\end{thebibliography}
\end{document}